# Preparative fractionation of a random copolymer (SAN) with respect to either chain length or chemical composition

Stefan Loske, Anja Schneider, and Bernhard A. Wolf*

Institut für Physikalische Chemie der Johannes Gutenberg-Universität Mainz and Materialwissenschaftliches Forschungszentrum der Universität Mainz, Welder-Weg 13, D-55099 Mainz, Germany

**ABSTRACT**

The possibilities to fractionate copolymers with respect to their chemical composition on a preparative scale by means of the establishment of liquid/liquid phase equilibria were studied for random copolymers of styrene and acrylonitrile (SAN). Experiments with solutions of SAN in toluene have shown that fractionation does in this quasi-binary system, where demixing results from marginal solvent quality, take place with respect to the chain length of the polymer only. On the other hand, if phase separation is induced by a second, chemically different polymer one can find conditions under which fractionation with respect to composition becomes dominant. This opportunity is documented for the quasi-ternary system DMAc/SAN/polystyrene, where the solvent dimethyl acetamide is completely miscible with both polymers. The theoretical reasons for the different fractionation mechanisms are discussed.

## Introduction

The *analytical* possibilities for the characterization of random copolymers with respect to their molecular and chemical non-uniformity are rather sophisticated[1-3]. This statement does, however, not hold true for the *preparative* separation of such macromolecules according to either their chain length or their chemical composition. Although homopolymers can be fractionated on an industrial scale by means of Continuous Polymer Fractionation[4] (CPF) or Continuous Spin Fractionation[5] (CSF), the application of these methods to copolymers is only at the beginning, despite the fact that their availability would be helpful for basic research as well as for certain industrial applications, where molecules containing too large a fraction of a certain comonomer may be harmful. In view of this situation we have studied the chances to develop a large scale fractionation technique for copolymers using the same procedures as for homopolymers. This constraint automatically excludes all copolymers containing monomers that form readily crystallizable sequences. Such a limitation appears however acceptable in view of the fact that there already exist preparative separation methods for these copolymers via fractional extraction[6]. Two other, more general options were also ruled out: The promising selective extraction by means of supercritical liquids[7] because of elaborate experimental requirements, and the use of demixing solvents[8] for similar reasons and in view of the limited efficiency expected from theoretical considerations as well as from orienting experiments[9].

The only examples for a successful *preparative* fractionation[10] of a copolymer was to our knowledge so far performed by means of CPF and concerns a copolymer made of carbonate and dimethylsiloxane. Due to the particularities of the system the separation takes place with respect to both parameters, molar mass and chemical composition, under certain conditions. The present work is dedicated to a more systematical exploration of the possibilities to tune liquid/liquid phase equilibria such that fractionation is predominately taking place according to one criterion only. To this end we have chosen two industrial samples of SAN, random copolymers of styrene and acrylonitrile, as the objects of investigation. With the sample of the lower chemical non-uniformity we are studying the possibilities offered by the marginal solvent toluene. According to literature reports[11] this system should be suitable for fractionation with respect to chemistry. In the past we have already studied solutions of SAN in toluene regarding the interfacial tension between the coexisting liquid phases formed upon the demixing[12]. By means of the SAN sample with the higher chemical non-uniformity we are investigating the possibilities, an induction of liquid/liquid phase separation by a properly chosen second polymer (instead of a low molecular weight solvent of marginal thermodynamic quality) offers for fractionation. In particular we wanted to know whether it is possible to find conditions under which the fractionation does predominantly take place with respect to the monomer content of the copolymer and not with respect to the molar mass *M*.

## Experimental Part

### *Materials*

Two different samples of poly(styrene-*ran*-acrylonitrile) (SAN) were used. Both are random copolymers and produced by BASF (Ludwigshafen) under the name Luran®.

One sample contains 40 mol% AN (25 wt%, elementary analysis) and exhibits a weight average molar mass $M_w$ of 147 kg/mol according to light scattering measurements in tetrahydrofurane (THF), it will be abbreviated as SAN 147w. Osmotic pressure measurements in THF yielded a number average of $M_n$ = 90 kg/mol. According to these data the molecular non-uniformity $U = (M_w/M_n) - 1$ is approximately 0.65. Because of the fact that the composition of this copolymer is identical with that of the azeotropic monomeric mixture, the chemical non-uniformity results exclusively from the differences in the monomer reactivity ratios. On the basis of theoretical considerations[13], one calculates an $e$ value of 0.1 (the parameter $e$ quantifies the chemical non-uniformity by analogy to $U$, measuring the molecular non-uniformity) from the monomer reactivity ratios tabulated in the polymer handbook[14].

The other SAN sample contains more AN, namely 48 mol% (32 wt%, elementary analysis) and has a $M_w$ value of 200 kg/mol (light scattering in DMAc). GPC measurements (PS-standards) yield an apparent molecular non-uniformity of $U^*$ = 1.2. This product is abbreviated as SAN 200w. The chemical non-uniformity calculated[13] for the instantaneously formed copolymer is on the same order of magnitude as for SAN 147w, namely 0.09. Due to the drift in the composition of the monomeric mixture associated with proceeding polymerization, the $e$ value of this technical product is necessarily considerably larger. More detailed information is unfortunately unavailable.

Polystyrene with narrow molecular weight distribution ($M_w$ = 1970 kg/mol according to light scattering in DMAc and $U$=0.1 from GPC) was purchased from PSS (Mainz); it is designated as PS 1970w.

Toluene (TL, puriss. quality) and N,N-Dimethylacetamide (DMAc, p.a. quality) were products from Fluka (Buchs, Switzerland). TL was dried over molecular sieves before use, DMAc was used as received.

*Phase diagrams*

**Binary system:** The cloud point curve of the binary system TL/SAN 147w was determined visually as described in literature[15]. Phase volume ratios as a function of composition for a constant depth of penetration into the two-phase regime of the phase diagram provided information on the critical conditions[16]. In order to obtain the tie lines for critical over-all composition, homogenous mixtures of critical composition were cooled down to different subcritical temperatures, where the liquid forms two phases. Macroscopic phase separation is typically reached within 1-2 days. After that the phases were separated by means of a syringe and weighted. The composition of these solutions was then determined by evaporating the solvent and weighting the remaining polymer. The chemical compositions of the so obtained polymer fractions were analyzed by NMR ($CDCl_3$ $^1H$, 200 MHz).

**Ternary system:** The cloud point curve of the ternary system was determined via titration: A solution of PS 1970 (15 wt%) in DMAc was added dropwise to different solutions of SAN 200w in DMAc (ranging in their concentration from 11 to 3 wt%) until the mixture becomes turbid. The composition at the cloud point was obtained by weighting. The method for the determination of the critical conditions was the same as with the binary systems[16]. To measure the tie lines of the ternary system, solutions were again titrated up to the cloud point, as described above, but then additional precipitant was added. Subsequently the resulting two-phase mixtures were heated until they became homogenous again and then cooled down slowly (1 K/min) to the desired final temperature. After macroscopic phase separation both phases were weighted and the solvent was removed by evaporation.

*Analysis*

The composition of the coexisting phases was determined by separating the solvent from the non-volatile components. The remaining polymer mixtures were then analyzed with respect to the weight fraction of the components by means of GPC (experimental details can be found in reference[17]). Due to the large differences in the molar masses of SAN and PS the corresponding peaks are sufficiently separated and can be used for analytical purposes. The calibration curve required for that purpose is shown in Figure 1.

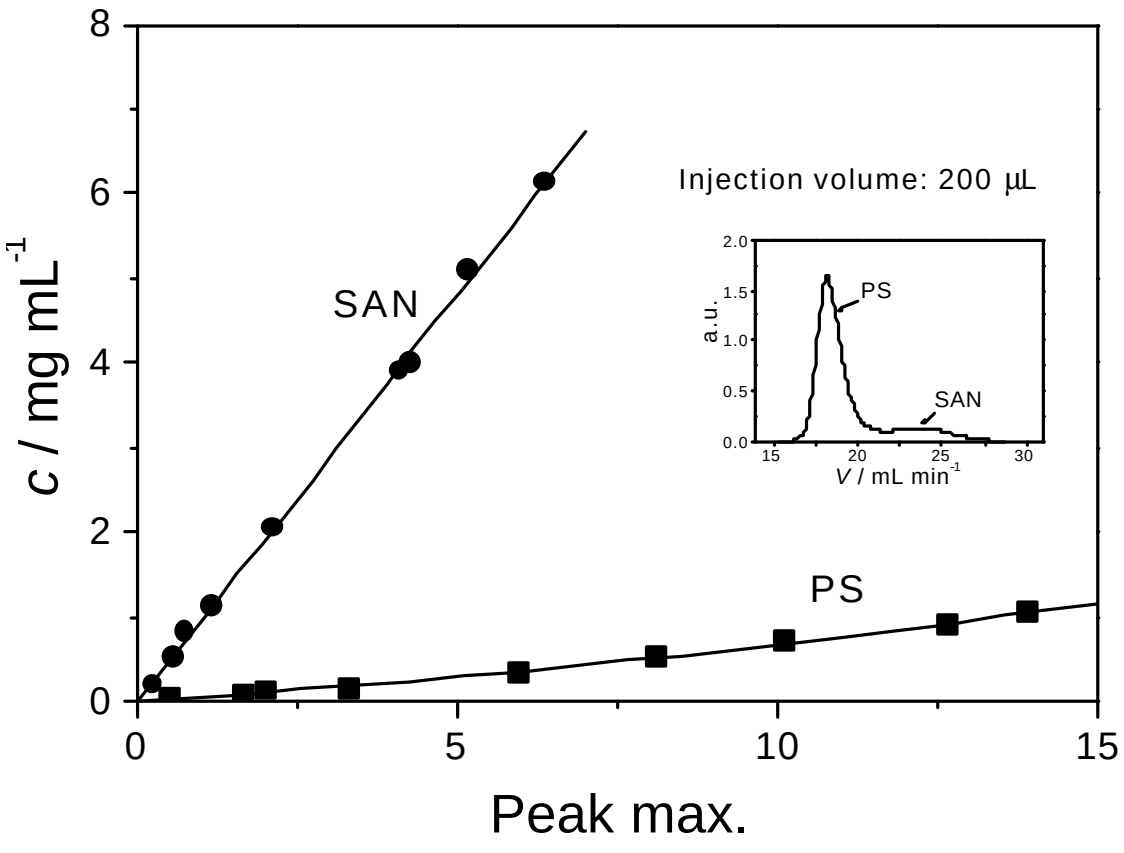

*Figure 1: Calibration curve to analyze mixtures of SAN and PS by means of GPC. For a constant injection volume of 200 L, the concentration of the individual components can be calculated from the measured peak maxima (cf. insert).*

The analysis of SAN with respect to its chemical composition was performed by means of a direct elementary analysis of the resulting blend. Despite the fact that the high molecular weight auxiliary polymer PS can be separated quantitatively from the SAN by titrating, either with methanol or acetone, this method was not applied because it would be associated with additional errors.

## Results and discussion

### *Fractionation by means of a low molecular solvent*

Figure 2 shows the phase diagram of the system TL/SAN 147w and Figure 3 an example for the fractionation with respect to the molar mass, namely for critical over-all composition of the mixture. In this case, and only under these conditions, one obtains a closed curve when plotting the average molar mass of the polymers contained in the coexisting phases as a function of temperature. The gel and the sol branch of the curve join at the critical temperature. The much larger difference between the molar mass of the sol fraction and that of the starting material, as compared with the difference of the gel fraction, reflects the asymmetry of the binodal line. This feature leads to an accumulation of most of the material in the gel phase, i.e. to little change in *M*, whereas only the shortest chains are contained in the sol phase.

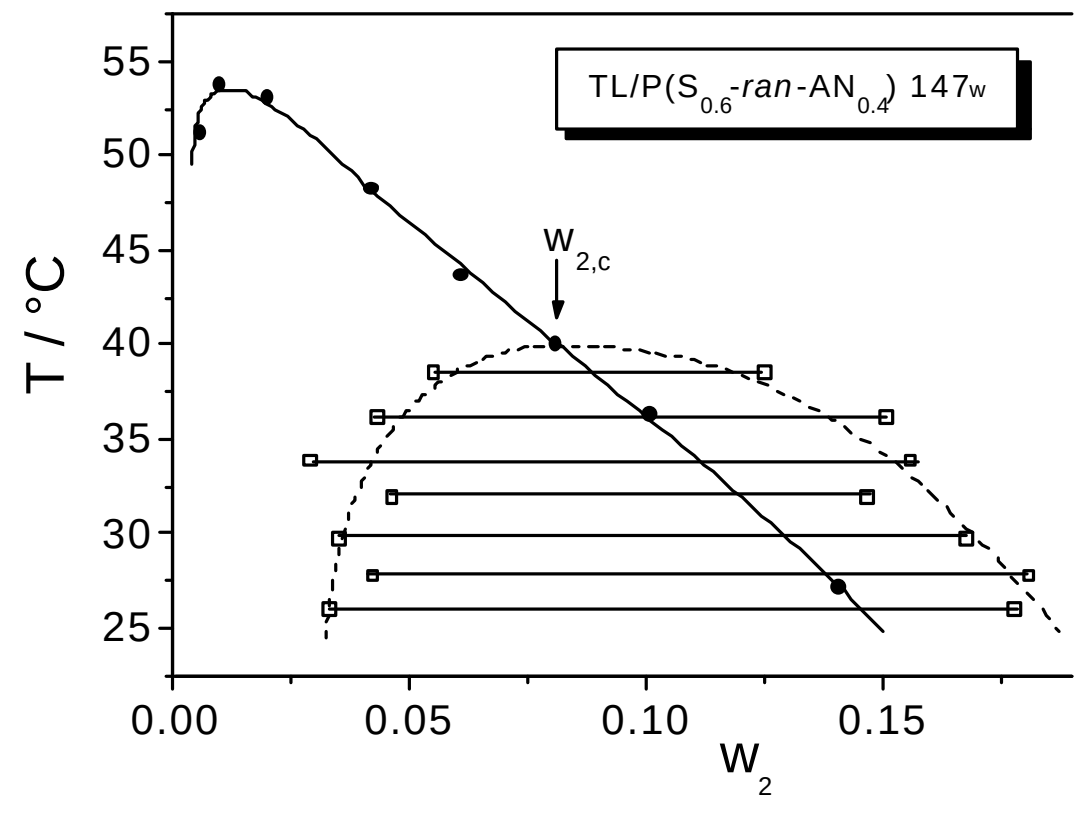

*Figure 2: Cloud points (full symbols) and cloud point curve (full line) of the system TL/SAN 147w. Also shown are some tie lines (open symbols) and the binodal line (broken) for the critical over-all composition the system.*

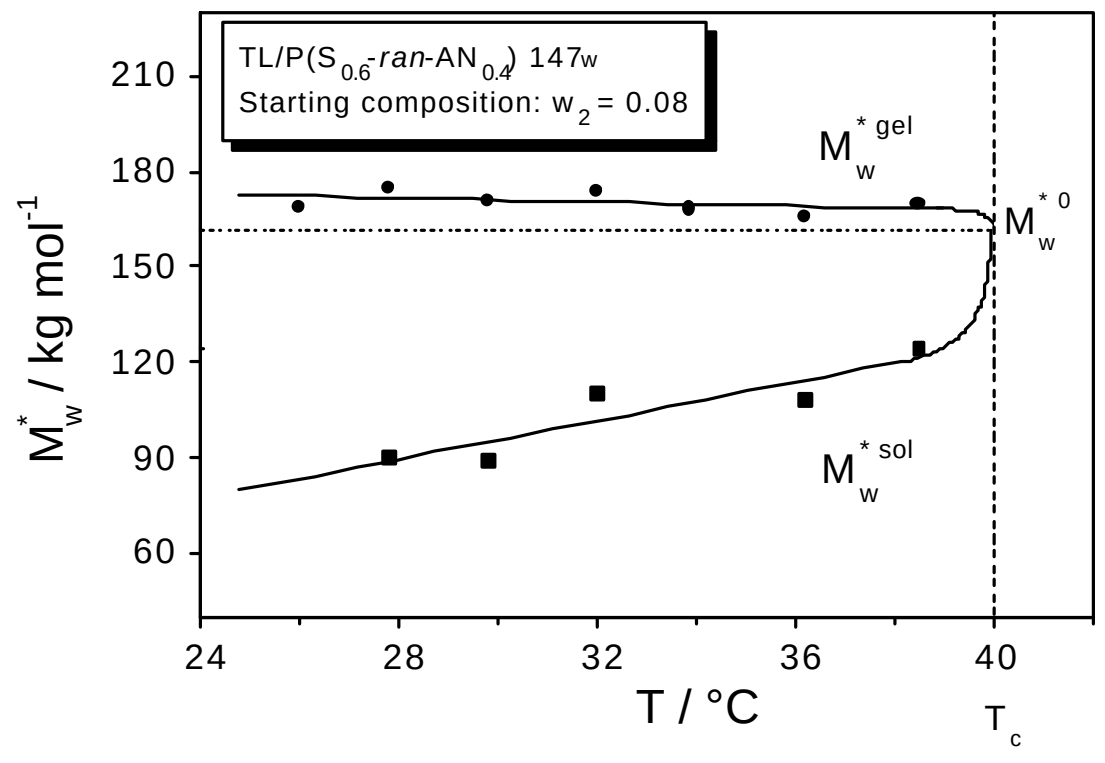

*Figure 3: Fractionation resulting from the partitioning of species differing in chain length on the coexisting phases (cf. Figure 2). The graph shows how the (apparent) weight average molar mass of the polymer contained in the polymer rich phase (gel) and in the polymer lean phase (sol) depends on temperature.*

In contrast to the results reported for the same system in the literature[11], the different fractions we have obtained in our experiments have all the same composition. The most likely explanation for this discrepancy lies in two features. In the first place we investigate an SAN sample with a comparatively narrow unimodal distribution, at variance with the previous study[11], where two such samples of different chemical composition were mixed (23 and 21 wt% AN). Secondly, we are for economic reasons checking the fractionation behavior at approximately ten times larger polymer concentrations. According to the present results a preparative fractionation of copolymers with respect to composition by means of marginal solvents does not appear very attractive because of the unsatisfactory separation efficiency at reasonably high polymer concentrations. In the next section we are therefore studying the fractionation that can be achieved by substituting the marginal solvent by an incompatible polymer.

*Fractionation by means of an auxiliary polymer*

The usual demixing of copolymer solutions into two liquid phases, greatly differing in their polymer content, does obviously not lead to an efficient separation of species with dissimilar chemical composition, as demonstrated in the last section. The pronounced differences in polymer concentrations favor the accumulation of the long chains in the gel (because of enthalpic reasons) and of short chains in the sol (because of entropic reasons) to such an extent that dissimilarities in chemical composition become secondary. For that reason we are here investigating whether the situation becomes better if the two phase state is not induced by unfavorable interactions between a high molecular weight compound and one (or several) low molecular weight component(s), but results from the incompatibility of chemically different polymers. Under these conditions the polymer concentration in coexisting phases can be kept comparable so that one can hope to eliminate or at least weaken the entropic driving forces to such an extent that enthalpic (i.e. chemical) effects become more efficient.

The idea to use a second polymer instead of the customary low molecular weight precipitant is not new. According to very early theoretical considerations[18] such systems could – at least in theory - be helpful for the fractionation of homopolymers. With regard to a practical realization of this concept the authors themselves were, however, rather skeptical because of minute differences in the densities of the coexisting phases and problems with the formation of emulsions. It is only recently that this polymer incompatibility could be successfully employed for a joint fractionation of two homopolymers[19], namely cellulose and PMMA. To our knowledge no attempts have so far been made to utilize the concept to fractionate copolymers with respect to their chemical composition.

For the fractionation of SAN we have chosen PS as the auxiliary component. In principle this pseudo-binary system, consisting of two polymer only, should already be apt for a separation with respect to chemical composition. However, its viscosity is even at elevated temperatures by far too high for that purpose. We therefore add DMAc as a low molecular weight compound to keep the viscosity of the mixtures manageable. The solvent is completely miscible with both polymers. The reason for selecting PS, instead of any other randomly chosen incompatible second polymer, is the perception that fractionation should be assisted if the auxiliary polymer is made of one of the monomeric species of the copolymer. Here we expect that the styrene rich copolymers will prefer the PS rich phase over the SAN rich phase.

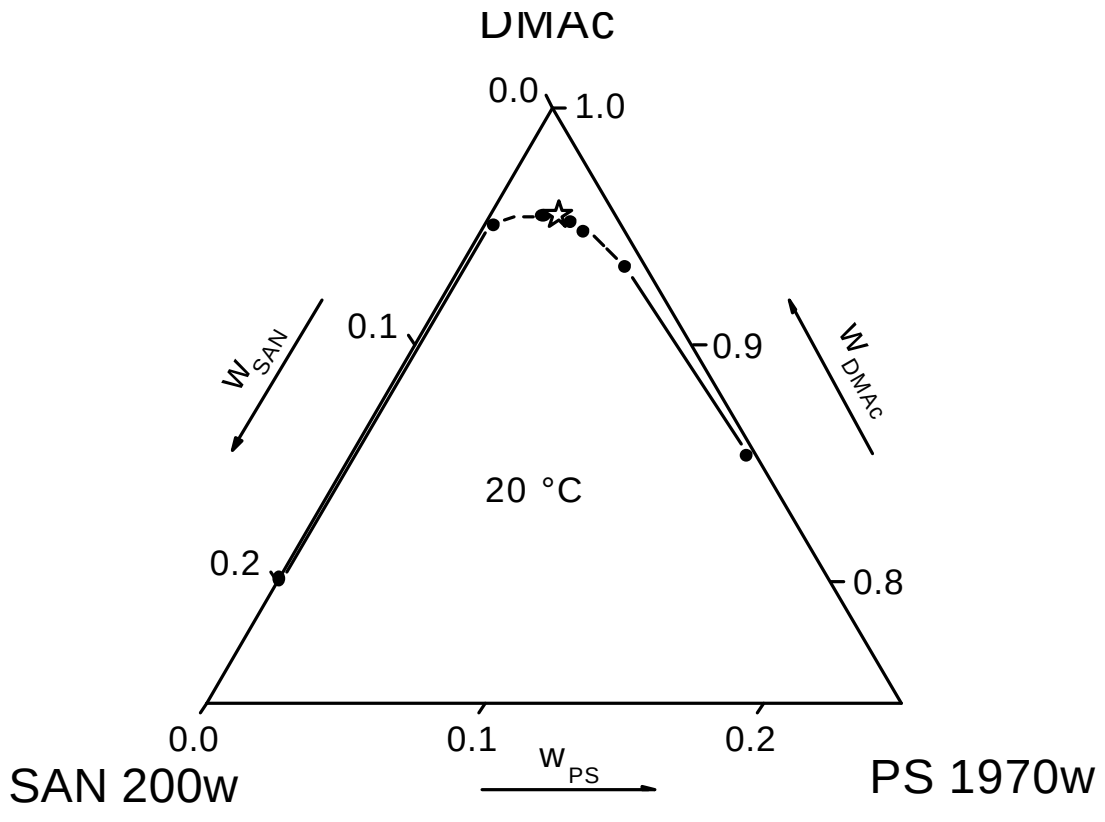

*Figure 4: Experimentally accessible part of the phase diagram of the system DMAc/SAN/PS at 20 °C. Full symbols: cloud points; star: critical point (2.3 wt% SAN and 2.4 wt% PS)*

As with all fractionation procedures it is mandatory to dispose of a certain minimum information on the phase diagram. The measured cloud point curve of the quasi-ternary system DMAc/SAN/PS and its critical composition at 20 °C are shown in Figure 4. In order to obtain a more comprehensive understanding of the present system, the phase diagram - above all within experimentally inaccessible regions - was also modeled, neglecting the polydispersities of SAN. For this purpose we have used a method developed by Horst[20]. It minimizes the Gibbs energy of mixing directly and does not require its derivatives. For these calculations the segment molar Gibbs energy of mixing $\overline{\overline{\Delta G}}$ of the ternary system is formulated as

$$\frac{\overline{\overline{\Delta G}}}{RT} = \frac{j_1}{N_1}\ln j_1 + \frac{j_X}{N_X}\ln j_X + \frac{j_Y}{N_Y}\ln j_Y + \\ + g_{1X} j_1 j_X + g_{XY} j_X j_Y + g_{1Y} j_1 j_Y \qquad (1)$$

where the segment size is arbitrarily fixed at 100 cm$^3$/mol. $N_i$ is the number of segments of species i. The index 1 stands for the solvent, X for PS and Y for SAN; $g_{1X}$, $g_{1Y}$, and $g_{XY}$ are the binary (integral) Flory-Huggins interaction parameters, which are for the present purposes assumed to be independent of composition. Furthermore the modeling does neither account for the molecular nor for the chemical non-uniformity of the copolymer. Despite these simplifications the calculated phase diagram should according to experience at least reflect some characteristic features.

For the interaction between homopolymer A and a random copolymer AB (here Y) the corresponding interaction parameters $g_{AY}$ and $g_{1Y}$ can be calculated[21] from *f*, the volume fraction of A units the copolymer contains and from the interaction parameter $g_{AB}$ between the segments of the homopolymers A and B as

$$g_{AY} = (1-f)^2 \, g_{AB} \qquad (2)$$

and

$$g_{1Y} = f\, g_{1A} + (1-f)\, g_{1B} - f\,(1-f)\, g_{AB} \qquad (3)$$

The interaction parameter for DMAc and PS was estimated from the second osmotic virial coefficient $A_2$ (a by-product of the light scattering experiments) according to

$$g \approx c = \left(\frac{1}{2} - A_2 V_1 r_2^2\right) \qquad (4)$$

where $V_1$ is the molar volume of the solvent and $r_2$ the density of the polymer. The parameters quantifying the interaction with polyacrylonitrile (PAN) $g_{(DMAc/PAN)}$ and $g_{(PAN/PS)}$ were adjusted such that the modeled binodal becomes as similar to the measured cloud point curve as possible. The calculations were performed with $f$ = 0.705 (following from the composition of the SAN sample), $g_{(DMAc/PS)}$ = 0.549, $g_{(DMAc/PAN)}$ = 0.482, and $g_{(PAN/PS)}$ = 0.034.

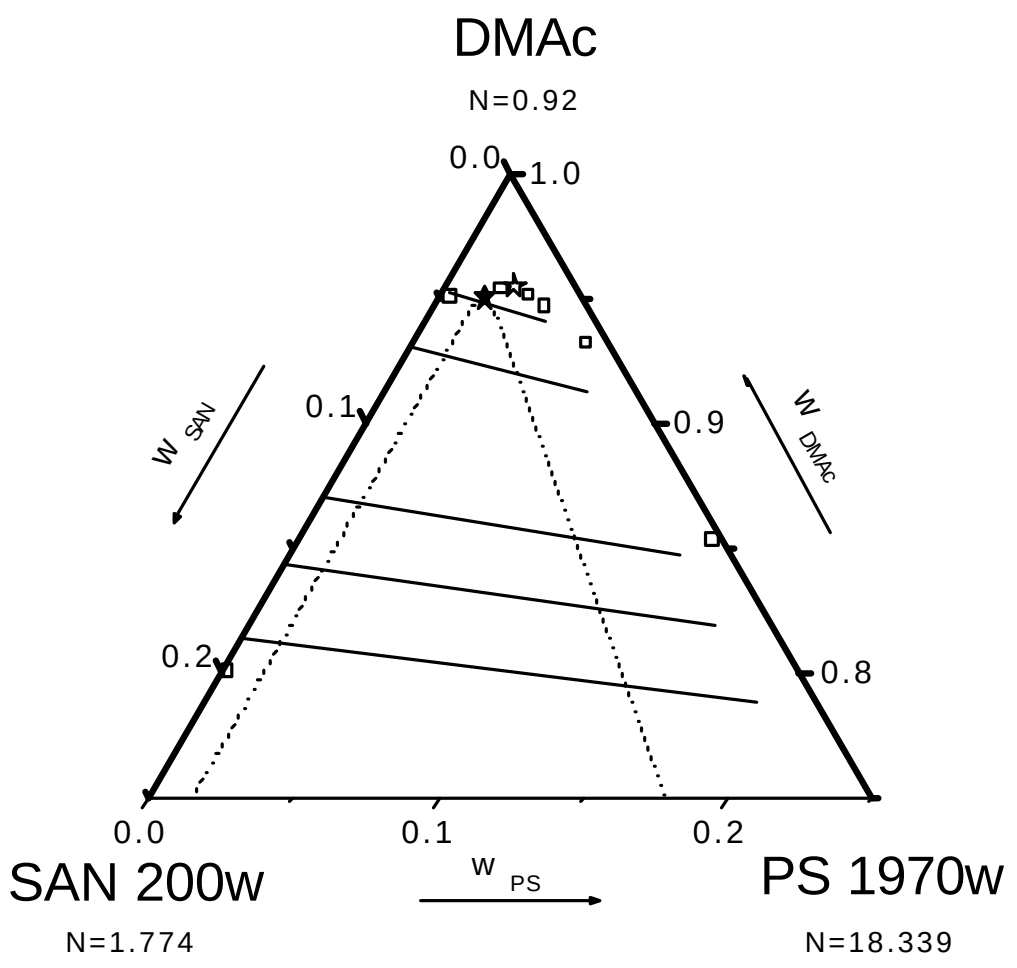

*Figure 5: Modeled phase diagram of the system DMAc/SAN/PS at 20 °C on the basis of the eqs (1) to (4), treating SAN as a single species. Individual points: Experimental data from Figure 4; full lines: calculated tie lines; dotted curve: calculated spinodal line.*

According to the calculated location of the critical point on the cloud point curve (approximately at its maximum, cf. Figure 5) and to the modeled position of the tie lines, the system should be suitable for the present purposes. The deviation of the tie lines from the desired parallelism with respect to the base line of the Gibbs phase triangle is inevitable and results from the large difference in the molar masses of the two incompatible polymers.

Figure 6 demonstrates that the tie lines of the real system, where SAN is molecularly as well as chemically rather non-uniform, deviates grossly from the calculations. The measured tie lines are considerably more tilted than the modeled ones. Presently it is unclear whether this is primarily so because of the non-uniformities, i.e. fractionation, or a deficiency in theory. With the interpretation of the experimental results it must in any case be taken into account that their illustration within a plane is inadequate. At least one additional dimension is required (e.g. perpendicular to the Gibbs phase triangle) to quantify the deviation in the composition of the copolymers contained in the coexisting phases from that of the original SAN. Only for the critical conditions of the system, where the tie lines degenerate into a point, two dimensions suffice. In the general case one end of the tie line is situated above the plane and the other below it.

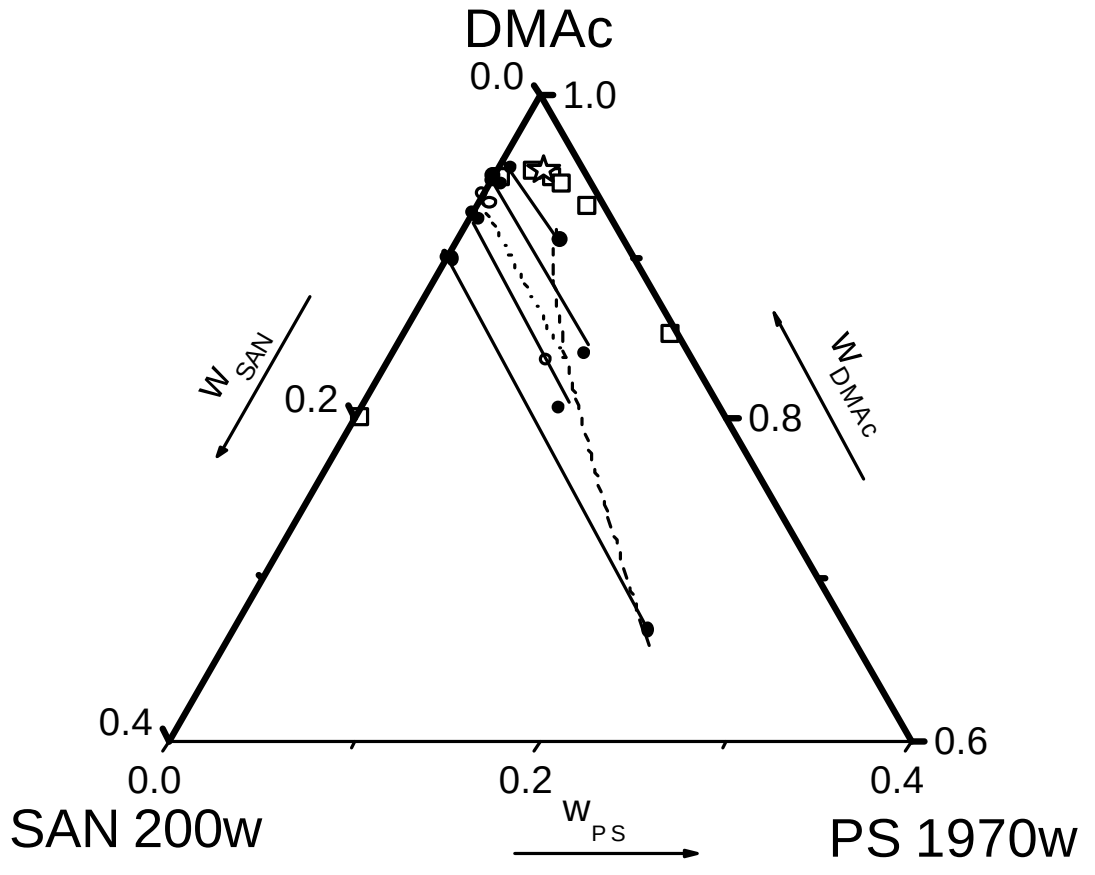

*Figure 6: Tie lines measured for the system DMAc/SAN/PS at 20 °C. The over-all composition of the different mixtures was in all cases chosen very close to the cloud point curve; due to experimental uncertainties in the determination of the composition of the PS rich phase the corresponding end of one tie line would in one case (open symbols) fall on another tie line, a situation which is thermodynamically impossible; for this reason its most likely position is represented by a dotted line. The broken line represents the PS rich branch of the binodal curve.*

We are now investigating by means of GPC experiments and elementary analysis to which extent fractionation with respect to molecular and chemical non-uniformity has taken place upon demixing under the present conditions. The GPC results of Figure 7 testify that the differences in the chain length of SAN between the original material and that contained in corresponding fractions remain minute. The maxima of the three curves are practically identical and the small displacement of the curves representing the fractions to lower $M$ values lies most likely within experimental error.

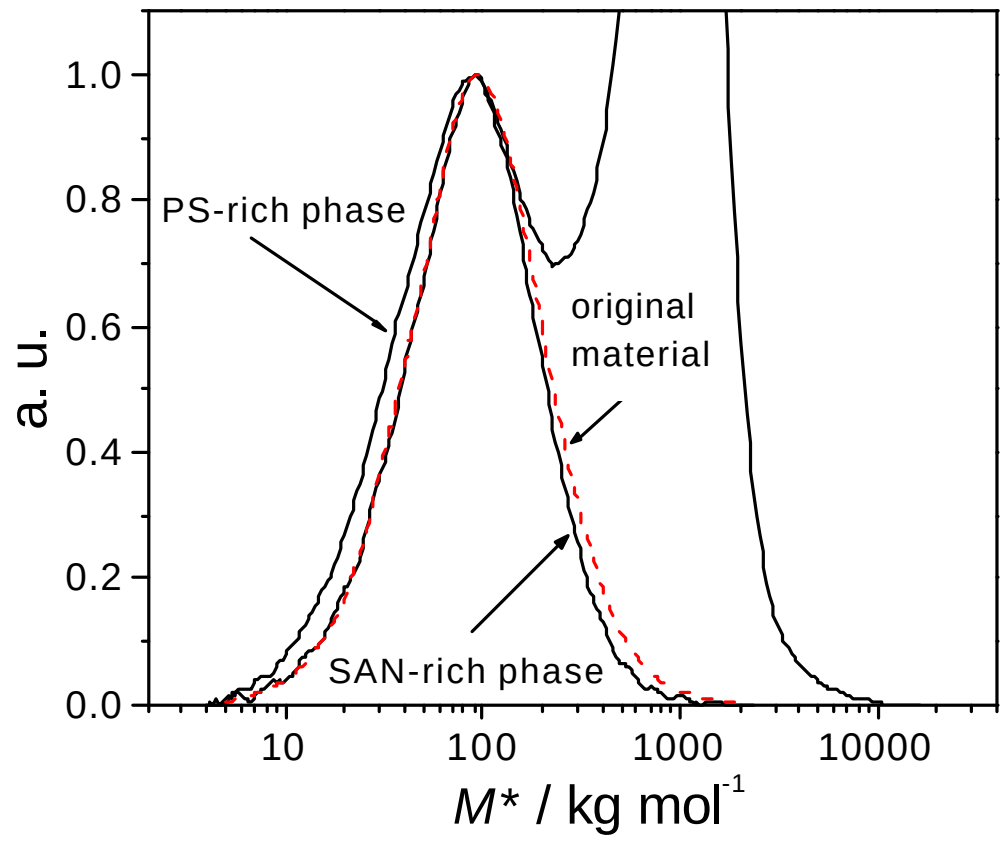

*Figure 7: Example for the GPC traces of the SAN fractions contained in the coexisting phases as compared with that of the unfractionated SAN.*

The chemical composition, on the other hand, varies markedly according to the data collected in **Tab. 1**. The styrene rich species of SAN are expectedly indeed accumulated in the PS-rich phase, whereas molecules rich in acrylonitrile are preferentially found in the SAN rich phase. From the present orienting experiments it is not yet possible to discern the optimum conditions for fractionation with respect to chemical composition. It may, however, be anticipated that they will lie not too far from the critical conditions, where the overall polymer concentrations in the coexisting phases are still comparable to damp the entropic driving forces and low enough to keep the viscosity of the mixture moderate.

## Conclusions

The orienting experiments concerning the possibilities to fractionate copolymers with respect to their chemical composition presented here show promise for the implementation of the large scale methods CPF and CSF mentioned in the introduction. According to the present results it should indeed be possible to use liquid/liquid phase equilibria to sort out chains that are harmful for special application (e.g. medical purposes or as additives in industrial processes) because they contain too large amounts of a certain monomer on a technical scale. It should go without saying that an optimization of these separation techniques for a certain task requires considerable further research. In this context one should keep in mind that favorable working points (over-all composition of the mixture) ought to be selected close to the cloud point curve and not too far from the critical composition of the system.

**Tab. 1.** Composition of SAN contained in the coexisting phases. Original composition of SAN 200w: 48 mol-% acrylonitrile. WP: working point = over-all composition

| | $w_{SAN}$ | $w_{PS}$ | **mol-% AN (SAN rich phase)** | **mol-% AN (PS rich phase)** |
|---|---|---|---|---|
| **WP** | 0.038 | 0.007 | 48 | 29 |
| **WP** | 0.048 | 0.007 | 49 | 31 |
| **WP** | 0.060 | 0.006 | 54 | 33 |
| **WP** | 0.071 | 0.006 | 56 | - |